# Order-disorder phase transition of cell membrane induced by THz irradiation studied via fluorescence recovery after photobleaching

Hiromichi Hoshina[*]

Center for Advanced Photonics, RIKEN, Sendai, Japan

hoshina@riken.jp

To elucidate the mechanism by which THz radiation non-thermally affects living organisms, the lateral diffusion constants of lipid molecules in the cell membranes of HeLa cells were measured using fluorescence recovery after photobleaching under THz wave irradiation (THz-FRAP) at frequencies of 0.10, 0.29, and 0.48 THz, with power densities ranging from 20 to 89 $mW/cm^2$. The potential heating effects of the THz irradiation were eliminated through temperature calibration using an ultrathin thermocouple, allowing for the investigation of the non-thermal effects of THz radiation. Irradiation at 0.10 and 0.29 THz induced an increase in diffusion constants at temperatures lower than the cell growth temperature. This suggests that THz irradiation induces the order-disorder phase transition of the cell membrane lipids by affecting the dynamics of bound water molecules. Our findings have important implications for the establishment of safety standards for THz radiation and for the potential development of new methods for cell manipulation using THz irradiation in the future.

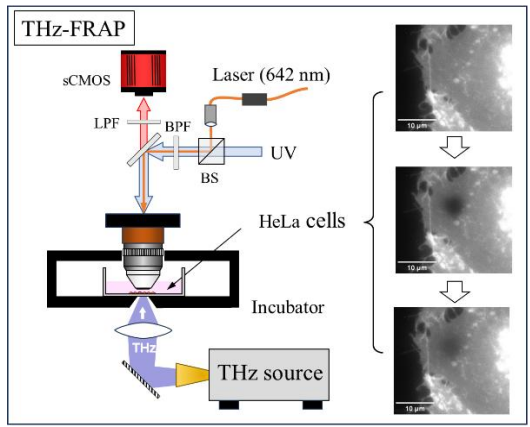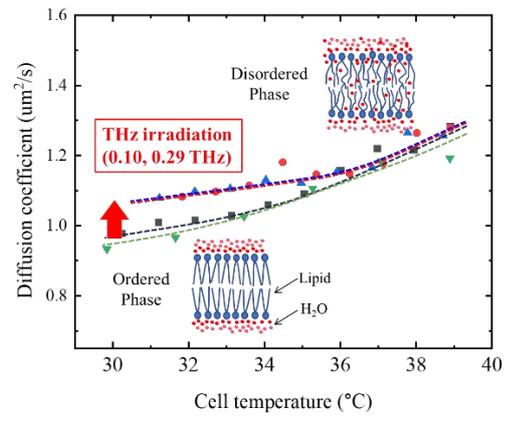

**Introduction**

Recent technological developments have enabled the development of various applications, such as motion sensors and next-generation wireless communications, using terahertz (THz) electromagnetic waves, with frequencies ranging from 0.1 to 10 THz.[1–5] Concurrently, there has been growing interest in the safety implications of THz electromagnetic waves, leading to numerous studies on the effects of THz wave irradiation on living cells.[6,7] For example, many researchers have investigated on the effects of irradiation on human skin tissues, because THz waves were incapable of penetrating deep into the human body due to the strong absorption by water molecules. Those results only show the thermal effects when the irradiation intensity is sufficient to significantly increase cell temperature, however, no remarkable changes was observed under lower power irradiaion.[8,9]

However, recent studies have demonstrated the non-thermal effects of THz radiation on certain living cells, organelles, and proteins. For example, Yamazaki et al. found that the reaction speed of actin filament formation is enhanced by exposure to 0.46 THz radiation.[10] This phenomenon was also confirmed in living HeLa cells, and it led to the inhibition of cell division during cytokinesis.[11] Non-thermal irradiation effects under weak THz radiation, which does not cause a temperature increase, have also been reported in studies on human fibroblasts[12,13], neuron cells[14,15] and mitochondrial membranes[16].

To elucidate the mechanisms underlying these findings, it is essential to understand how the energy of THz photons induce dynamic and structural changes in biomolecules and surrounding water. THz waves absorbed by living organisms excite the collective motion of biomolecules and surrounding water molecules,[17–19] potentially inducing conformational changes in biomolecules before energy relaxes to the thermal equilibrium. In a recent study, Sugiyama et al. observed non-thermal acceleration of protein hydration under 0.1 THz irradiation.[20] They investigated an aqueous lysozyme solution, which is not thermally equilibrated, using dielectric relaxation and nuclear magnetic resonance techniques. Their results indicate that sub-THz irradiation excites the dynamics of hydrated water and helps reconstruct the hydrogen bond network around proteins, facilitating the entry of water molecules into hydrophobic cavities.

This study aimed to investigate the irradiation effect of THz wave on the dynamics of living cell membrane. The hydrophilic groups on the outer surface of the lipid bilayer form strong hydrogen bonds with surrounding water molecules. By analogy with Sugiyama et al.'s findings,[20] THz radiation may alter the dynamics of water molecules bound to lipid bilayers, which could in turn affect the properties of cell membranes. Non-thermal effects of THz irradiation have been reported to induce changes in membrane functions in living cells.[14–16] To quantitatively evaluate changes in cell membrane dynamics, the lateral diffusion of membrane molecules in living HeLa cells was measured using fluorescence recovery after photobleaching (FRAP). FRAP is a widely used method for observing lateral diffusion characteristics of cell membranes.[21] We developed a microscope capable of performing FRAP during THz wave irradiation(THz-FRAP) and used it to measure the diffusion constants of the plasma membranes of living HeLa cells.

**FRAP measurement of HeLa cells with THz irradiation**

Fig. 1(a) presents typical results from the FRAP measurements (see supplemental information). The bleached fluorophores in the cell membrane produce a dark spot in the microscope image (post-bleach). Fig. 1(b) shows the cross-section of the intensity difference between the pre-bleach and post-bleach images. The size of the bleached spot is estimated using least squares fitting with a Gaussian function, and its full width at half maximum (FWHM) is used to determine the region of interest (ROI) diameter (approximately 6 mm). The dark spot gradually dissipates over time due to the lateral diffusion of membrane molecules (after diffusion).

Fig. 1(c) plots the average fluorescence intensity of the ROI and a control area. The ROI intensity was normalized by that of the control area, which slightly diminishes over time due to UV light exposure, and fitted with an exponential function. The obtained recovery rate, $\tau$, was used to calculate the lateral diffusion constant, $D$, according to the equation $D = \omega^2/4\tau$, where $\omega$ is the radius of the ROI. For example, the result shown in Fig. 1 yields a diffusion constant of $D = 1.97$ mm$^2$/s.

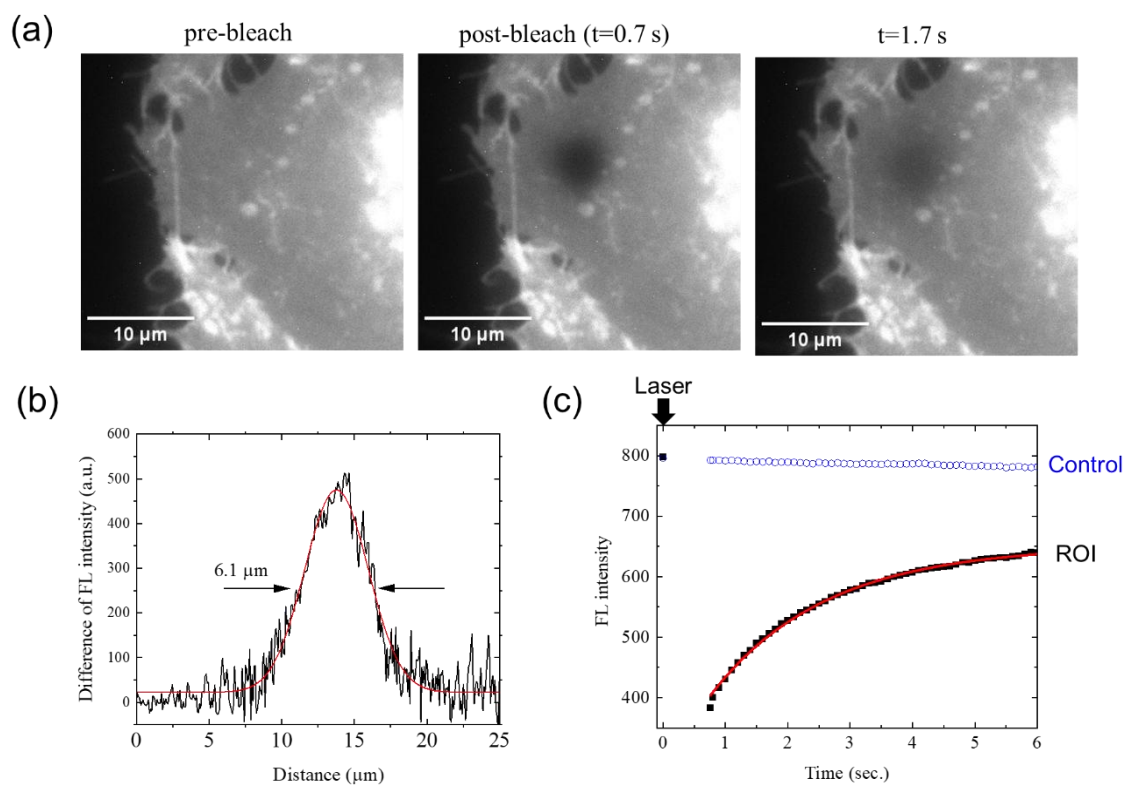

Fig. 1 Typical results from the FRAP measurements. (a) Fluorescence images of HeLa cells at t = −1 s, t = 0.7 s, and t = 1.7 s, respectively, where the excitation laser was applied for 100 ms at t = 0 s. (b) Cross-section of the intensity difference between the pre-bleach and post-bleach images (black), along with the result of least squares fitting using a Gaussian line function (red). (c) The average fluorescence intensity of the ROI (black) and the control area (blue), with the result of least squares fitting using an exponential function.

Fig. 2 shows the lateral diffusion coefficient (D) measured as a function of the incubator temperature, $T_1$, under various conditions: (a) without THz irradiation, (b) with 0.10 THz irradiation, (c) with 0.29 THz irradiation, and (d) with 0.48 THz irradiation. In all graphs, the average diffusion coefficient increases with temperature. To compare the data at equivalent sample temperatures, $T_s$, the effect of temperature increases due to THz radiation was corrected using calibration data obtained with an ultrathin thermocouple. Fig. 3 shows the mean values and standard deviations of the diffusion coefficients as a function of sample temperature. A clear difference was observed between the data for 0.10 THz and 0.29 THz irradiation and the control data (no THz irradiation) below 36 °C. In contrast, no significant difference was observed between 0.48 THz irradiation and the control. The box plots in the inset compare the diffusion coefficients at the closest sample temperatures. A t-test analysis revealed statistically significant differences between the 0.10/0.29 THz groups and the control, whereas no significant difference was observed between the 0.48 THz irradiation group and the control sample.

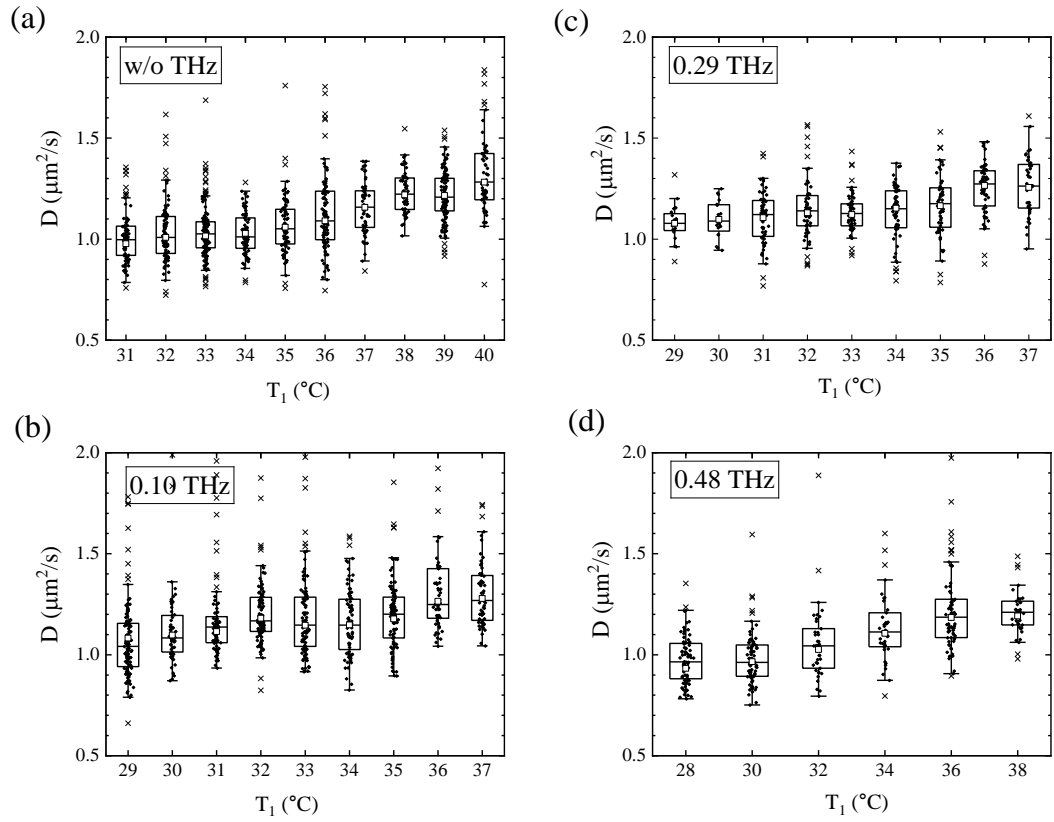

Fig. 2 Lateral diffusion coefficients (D) of the HeLa cell membranes measured at various incubator temperatures, $T_1$, under different conditions: (a) without THz irradiation, (b) with 0.10 THz irradiation, (c) with 0.29 THz irradiation, and (d) with 0.48 THz irradiation. Open squares represent the average values and crosses indicate outliers.

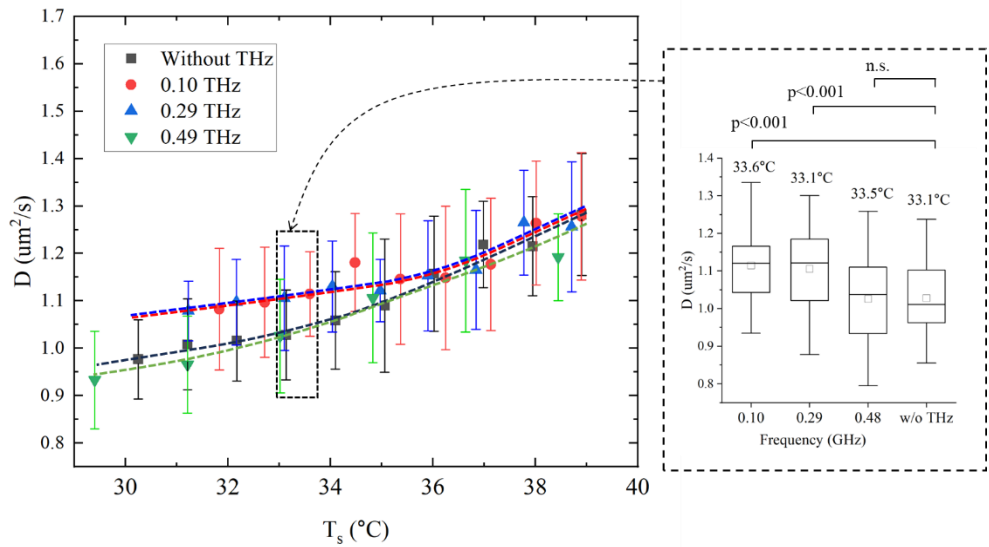

Fig. 3 Mean values of the lateral diffusion coefficients of HeLa cell membranes are plotted with cell temperature $T_s$. Error bars represent the standard deviations, and the dashed lines are intended as eye guides. The inset shows box plots and mean values (open squares) comparing the diffusion coefficients measured at similar sample temperatures, i.e., with THz irradiation of 0.10 THz ($T_s$ = 33.6 °C), 0.29 THz ($T_s$ = 33.1 °C), 0.48 THz ($T_s$ = 33.5 °C), and without THz irradiation. The p-values above the box plots indicate the results of the t-test.

**Discussion**

Differential scanning calorimetry (DSC) reveals that lipid bilayers undergo phase transitions with temperature, which was observed not only in artificially created lipid membranes but also in living cells.[22] For example, Heimburg et al. measured the heat capacity profiles of *E. coli* bacteria using DSC and identified a broad phase transition occurring a few degrees below the cell growth temperature.[23] Farber et al. observed the phase transition of the HeLa cell membrane through Laurdan fluorescence spectroscopy.[24] The phase transition in cell membrane affects various physical properties such as heat capacity, adhesion, permeability,[25] and diffusion. Therefore, the transition from a disorder to an ordered phase can change cellular function when the temperature is reduced. To adapt to environmental changes, cells modify their membrane composition to shift the transition toward lower temperatures when cultured in cooler environments.[22]

Fig. 4 illustrates the schematic of the phase transition of the cell membrane, showing heat capacity, cell growth temperature,[23] and the diffusion coefficients observed in this study (Fig. 3). When the sample temperature falls below the cell growth temperature, the fraction of ordered lipid membrane increases as the temperature decreases, reflected in the reduction of lateral diffusion coefficients. However, irradiation with 0.10 and 0.29 THz radiation increases the diffusion coefficients at lower temperatures. This effect is particularly pronounced below the cell growth temperature of 37 °C, suggesting that THz irradiation "melts" the ordered structure of the cell membrane.

Recently, Leung et al. observed a correlation between the $^2$H NMR order parameter and the generalized polarization of the fluorescence probe Laurdan,[26] concluding that the order of lipid molecules is closely linked to the quantity and dynamics of water molecules at the glycerol backbone level of the membrane. This finding strongly suggests that the induction of the order-to-disorder transition by THz radiation is also influenced by the behavior of water molecules surrounding the lipids. THz irradiation at 0.10 and 0.29 THz may excite the dynamics of these water molecules, promoting their entry into the hydrophobic regions of the lipids.

As reported by Sugiyama et al., sub-THz irradiation excites the hydration water around biomolecules, facilitating their entry into hydrophobic cavities.[20] In this study, the irradiation effect of 0.1 and 0.29 THz on the cell membrane appears to similarly excite the dynamics of hydration water around the hydrophilic groups of lipids, allowing water to penetrate the hydrophilic regions. As a result, the cell membrane undergoes an order-to-disorder transition. This phenomenon, however, was not observed with 0.48 THz radiation, likely for two reasons. First, the intensity of THz radiation from the 0.48 THz source is lower than that from the other two sources, as shown in Table 1. Second, the excitation modes of hydrated water molecules exhibit frequency dependence. Yamamoto et al. reported that the relaxation modes of hydrated purple membrane occur at lower (sub-THz) frequencies whereas vibrational modes dominate at higher frequencies (>1 THz).[27] If the observed effects are due to the excitation of these relaxation modes, the excitation efficiency decreases at higher frequencies.

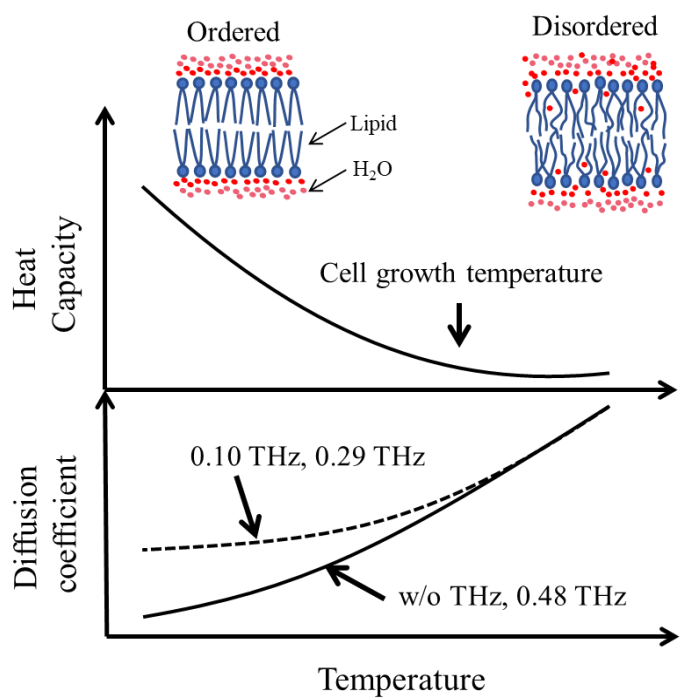

Fig. 4 Schematic representation of the phase transition, including heat capacity, cell growth temperature,[23] and diffusion coefficients observed in this study (Fig. 3).

**Conclusion**

In this study, the lateral diffusion constants of lipid cell membranes in HeLa cells were measured using FRAP under cw-THz wave irradiation at 0.10, 0.29, and 0.48 THz. To eliminate the effects of the heating caused by THz irradiation, the sample temperature during measurements was calibrated using an ultrathin thermocouple. Under THz irradiation at 0.10 and 0.29 THz, the diffusion constants of lipid molecules were non-thermally increased when the sample temperature was lower than the cell growth temperature, a condition where the ordered phase of the cell membrane is predominant. These findings suggest that THz irradiation disrupts the ordered phase of the cell membrane in a non-thermal manner. We believe that our study represents a significant step toward elucidating the mechanism by which THz radiation non-thermally affects living organisms. Our findings have important implications for the establishment of safety standards for THz radiation and for the potential development of new methods for cell manipulation using THz irradiation in the future.

**Methods**

**Experimental setup for THz-FRAP measurement**

Fig. 5 (a) illustrates the experimental setup for the THz-FRAP measurement. HeLa cells were cultured in a film-bottom dish containing 2 ml of cell growth medium, which was placed in an incubator for 3 days at 37°C. The samples were observed using a 60× water immersion lens (Olympus, LUMFLN60XW). The sample temperature was regulated by a heater located at the bottom of the incubator (T = $T_1$) and an objective lens heater (T = $T_1$ + 1.0 °C). Broadband UV light (Thorlabs, X-Cite 200DC lamp) for fluorescence imaging and laser light for bleaching (642 nm, 40 mW, Thorlabs, S4FC642) were overlayed using a cube beamsplitter and directed to the objective lens through a bandpass filter (BPF) (635–645 nm) and a dichroic mirror (650 nm). The laser light was introduced through a single-mode optical fiber (mode field diameter = 4.6 μm) combined with a fiber collimator (Thorlabs F230APC-633) and a variable optical attenuator (VOA; Thorlabs, V600F) to achieve confocal alignment at the sample plane. The laser power at the sample position was set to 5 mW. The fluorescence image of the sample was monitored using an sCMOS camera (Thorlabs, CS2100M-USB) equipped with a low-pass filter (LPF) (675 nm), and the images were recorded on a PC.

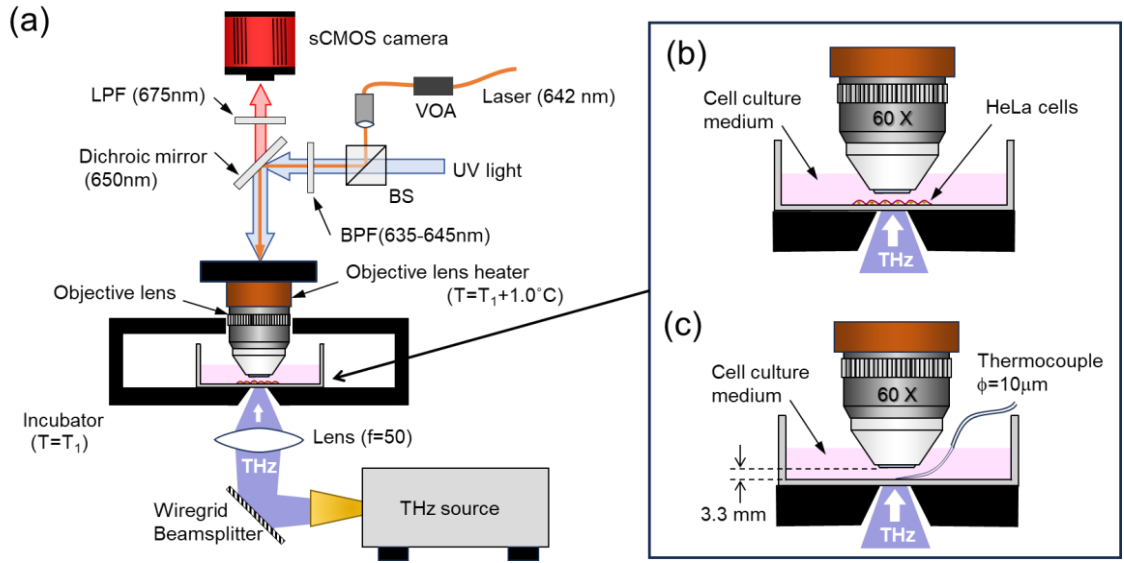

Fig. 5 (a) Experimental setup for THz-FRAP. (b) THz irradiation setup for HeLa cells. (c) Calibration setup for measuring sample temperature during THz irradiation.

**THz irradiation**

The THz beam was irradiated from the bottom of the incubator through a 5-mm aperture. The output of the THz source was reflected by a wire grid beamsplitter and focused by a convex lens with a 50-mm focal length made by Tsurupica® (PACS Inc.). Three types of cw THz sources were used: 0.10 THz and 0.29 THz (TeraSense Inc.), and 0.48 THz (ACST GmbH). The specifications of the THz sources are summarized in Table 1. The THz beam profile and power density under each irradiation condition were measured using a power meter (THz20, SLT Sensor- und Lasertechnik GmbH) combined with a 0.5-mm-diameter aperture. The beam profile was fitted to a Gaussian function, and the power density at the sample position was estimated. The sample was placed at the focal point of the THz beam, where the beam size (~5 mm) was significantly larger than the observation area of the microscope (~100 μm). Consequently, the THz power density could be considered uniform, with a variation of less than 10%. The estimated power density of the sample is listed in Table 1.

Since HeLa cells adhere to the bottom of the dish, the THz beam directly irradiated the cells without absorption by the culture medium (Fig. 5(b)). The absorption of the film-bottom dish was minimal, less than 5%. Assuming that the absorbance of HeLa cells in the THz region is similar to that of liquid water, which is 37, 60, and 75 cm$^{-1}$ at 0.10, 0.29, and 0.48 THz, respectively[28], the penetration depths of the THz radiation were 117, 73, and 58 μm, respectively. which exceed the thickness of HeLa cells (10−30 μm), ensuring that the entire cell region was exposed to THz radiation. (see Supplemental information)

| Frequency (THz) | Source power (mW) | FWHM (mm) | Power density at the sample position (mW/cm²) | Temperature change of sample ($\Delta T_s$) (°C) |
|---|---|---|---|---|
| 0.10 | 45 | 5.02 | 89 | 2.5–3.0 |
| 0.29 | 30 | 4.57 | 65 | 2.3–2.6 |
| 0.48 | 12 | 5.68 | 20 | 2.4–2.8 |

Table 1. Specifications of the THz sources, beam spot size (FWHM), irradiated power density at the sample position, and temperature change of the sample during irradiation (see also Supplemental information).

**Calibration of cell temperature**

During THz irradiation, the sample temperature increases due to the absorption of THz photon energy. To exclude the thermal effect from the measurements, the sample temperature during irradiation was carefully calibrated. To calibrate the sample temperature, $T_s$, the temperature at the bottom of the dish was measured by using ultra-fine K-type thermocouple with 13-μm thickness (KFG-13, ANBE SMT Co.), and a calibration curve was generated as a function of $T_1$ (Fig. 5(c)). The temperature change of the sample during the irradiation is shown in Table 1.

**Sample preparation**

HeLa cells (provided by Prof. Masahiko Harata at Tohoku university) were seeded on a 0.15-mm-thick film-bottom dish and cultured in Dulbecco's Modified Eagle Medium (Gibco) supplemented with 10% fetal bovine serum and antibiotics (penicillin and streptomycin) at 37 °C in a 5% $CO_2$ humidified atmosphere for 2 d. The cell membranes were stained in the culture medium for 30 min using 0.5% CellBrite® red cytoplasmic membrane dyes. After staining, the cells were rinsed three times with fresh, warm growth medium and incubated at 37 °C for two additional 5-min intervals.

**FRAP measurement**

The FRAP measurement was conducted as follows: First, the laser was irradiated for 100 ms at t = 0, and fluorescence images were captured every 100 ms after a delay of 700 ms to prevent phosphorescence. The image resolution was 1920×1024 pixels with a scale of 0.075 μm/pixel. The recorded images were stored on a PC and analyzed using LabVIEW™ software. Initially, the position and size of the ROI were determined by fitting the intensity of the bleached area with a Gaussian function. Fuorescence intensities within the ROI were then measured. The obtained intensities were fitted to an exponential decay function, and the decay rate τ was calculated.

FRAP data were collected for each sample by alternating the THz irradiation between ON and OFF. Measurements were repeated with more than four samples under each condition.

The order of measurements, with and without THz irradiation, was reversed for each sample. For example, 20 points were measured without THz, followed by 20 points with THz, and vice versa.

**Acknowledgments**

This work was supported by Grant-in-Aid for Scientific Research (KAKENHI) No. 22K18992 and 23H01859 from the Japan Society for Promotion of Science (JSPS). We would like to express our gratitude Prof. Masahiko Harata (Tohoku University) for providing Hela cells and having insightful discussions with authors. We also thank Dr. Keiichiro Shiraga (Kyoto University) for kindly providing absorbance data of water in THz frequency region. We would like to express our gratitude to the Terahertz Sensing and Imaging Team, RIKEN for their support for the experiment.

**Data availability**

The datasets generated and/or analyzed during the current study are not publicly available due to the size of data but are available from the corresponding author on reasonable request.